\documentclass[prl,twocolumn]{revtex4}
\usepackage{graphics}
\usepackage{graphicx}
\usepackage{amsmath}
\usepackage{amstext}
\usepackage{xcolor}
\usepackage{}
\usepackage{amssymb}
\usepackage{hyperref}

\newcommand{\ex}[1]{\mathrm{e}^{#1}} 

\newcommand{\im}{\mathrm{i}}
 
\newcommand{\bos}[1]{\boldsymbol{#1}}
\newcommand{\vect}[1]{\boldsymbol{#1}}

\begin{document}
\title{Hidden vortex lattices in a thermally paired superfluid}
\author{E. K. Dahl$^{1}$, E. Babaev$^{2,3}$ and A. Sudb{\o}$^{1}$}
\affiliation{
$^{1}$Department of Physics, Norwegian University of Science
and Technology, N-7491 Trondheim, Norway\\
$^{2}$Physics Department, University of Massachusetts, Amherst MA 01003, USA\\
$^{3}$Department of Theoretical Physics, The Royal Institute of Technology 10691 Stockholm, Sweden}

\date{\today }
\begin{abstract}
We study the evolution of rotational response of a hydrodynamic model of a
two-component superfluid with a non-dissipative drag interaction, 
as the system undergoes a transition into  a paired phase at finite temperature.
The transition manifests itself in a change of (i) vortex lattice symmetry, and (ii) 
nature of vortex state. Instead of a vortex lattice, the system forms a highly disordered 
tangle which constantly undergoes merger and reconnecting processes involving different 
types of vortices, with  a ``hidden" breakdown of translational symmetry.
\end{abstract}
\maketitle

Recently, there has been increased interest in so-called ``paired" states of superfluids 
(and also related counter-flow states) where pairing results from proliferation of 
{\it composite} topological defects in various physical contexts \cite{PSF1}-\cite{Dahl}. 
The mechanism can be outlined as follows. In certain systems
the energetically cheapest defects which proliferate  under the influence of thermal 
fluctuations or applied external field are not the simplest vortex loops.  Rather, they  are 
composite ones. That is,  they have phase winding in several components of the order parameter, 
but nonetheless lack topological charge in some sector of the model. Consequently, their 
proliferation does not restore symmetry completely.  Broken symmetry may remain in e.g. 
sums of the phases of components of the order parameter, and the resulting state is frequently 
called a paired superfluid. Since the origin of pairing in this case is an entropy-driven 
formation of a tangle of composite topological defects, one encounters an unusual situation 
in which a system forms paired states as a consequence of heating. Thus, in what follows we will 
refer to this state as a Thermally Paired Superfluid (TPS), to distinguish it from a 
conventional pairing mechanism.
\par
Today, the experimentally most feasible system to study TPS, appears to be the multicomponent 
Bose--Einstein Condensates. Here, TPS can arise \cite{PSF2,Dahl} due to a current-current 
(Andreev--Bashkin) interaction, which can be tuned in an especially wide range for bosons in optical lattices 
\cite{PSF2}. Questions therefore arise as to how the transition into a TPS alters the rotational response 
of the system, and what are its experimental signatures? In this paper, we address this by studying 
a hydrodynamic model of a mixture of two superfluids with a dissipationless drag \cite{PSF2,Dahl}, given 
by \begin{align}
\label{model}
{\it F}=\frac{1}{2}
%\int\!\du\bos{r}
\bigg\{\sum_{i=1,2}&m_i n_i\left(\frac{\hbar\nabla\theta_i}{m_i}-\vect{\Theta}\right)^2\nonumber \\
%&+m_2\rho_2\left(\frac{\hbar\nabla\theta_2}{m_2}-\bos{r}\times\bos{\Omega}\right)^2\nonumber \\
&-\sqrt{m_1m_2} n_d\left(\frac{\hbar\nabla\theta_1}{m_1}-\frac{\hbar\nabla\theta_2}{m_2}\right)^2\bigg\},
\end{align}
where $\theta_i$, $n_i$ and $m_i$  are the phases, densities  and masses of the condensates respectively, 
the angular frequency is given by $\bos{\Omega}=\nabla\times\vect{\Theta}$, while $ n_d$ is the density of 
one component dragged by the other. The central feature of the model  is that 
for significantly strong drag $n_d$ the  composite vortices with phase winding in both components 
$(\Delta \theta_1=2\pi,\Delta \theta_2=-2\pi)$ [in what follows denoted by $(1,-1)$], become the  
energetically cheapest to excite and are the easiest objects of a thermal fluctuation-driven 
proliferation  \cite{PSF2,Dahl}. The resulting phase is well described by separating out the sector of 
the model unaffected by proliferation of composite vortices. The accuracy of this procedure 
was numerically checked in various regimes in Ref. \onlinecite{Dahl}. The Hamiltonian reads, 
after separation of variables
\begin{align}
{\it F}=\frac{1}{2}\bigg\{&
\left(
\frac{\frac{n_1n_2}{m_1m_2}-\frac{n_d}{\sqrt{m_1m_2}}\left(\frac{n_1}{m_2}+\frac{n_2}{m_1}\right)}
   {\frac{n_1}{m_1}+\frac{n_2}{m_2}-
%\frac{
\sqrt{m_1m_2}
%}
(m_1^{-1}+m_2^{-1})^2 n_d}
\right)\times\nonumber \\
&\left(
\hbar\nabla\theta_1+\hbar\nabla\theta_2-\left(m_1+m_2\right)\vect{\Theta}
\right)^2\nonumber \\
&+
%\frac{1}
\left[ \frac{n_1}{m_1}+\frac{n_2}{m_2}-\sqrt{m_1m_2}(m_1^{-1}+m_2^{-1})^2 n_d \right]^{-1}
\times \nonumber \\
&\left[
\left( \frac{n_1-n_d\frac{m_1+m_2}{\sqrt{m_1m_2}}}{m_1}
\right)
\left( \hbar\nabla\theta_1-m_1\vect{\Theta}
\right) \right.\nonumber\\
&\left.-
\left( \frac{n_2-n_d\frac{m_1+m_2}{\sqrt{m_1m_2}}}{m_2}
\right)
\left( \hbar\nabla\theta_2-m_2\vect{\Theta}
\right)
\right]^2,
\end{align}
where the first term represents the part of the model unaffected by proliferation of $(1,-1)$ vortices.
In what follows, we consider the case  $m_1=m_2=1$ and $n_1=n_2=\rho$,  $n_d=\rho_d$,  
in units where $\hbar=1$, namely
 \begin{align}
\label{compmodel}
F%=&\frac{1}{2}
%\int\!\du\bos{r}
%\bigg\{
%\frac{n}{m}\left(\left(\nabla\theta_1-m\vect{A}\right)^2 +\left(\nabla\theta_2-m\vect{A}\right)^2\right)-\frac{n_d}{m}\left(\nabla\theta_1-\nabla\theta_2\right)^2
%\bigg\}\nonumber \\
=%\frac{1}{2}
%\int\!\du\bos{r}
%\bigg\{
\frac{\rho}{4}\left(\nabla(\theta_1+\theta_2)-2\vect{\Theta}\right)^2 %\nonumber \\
+\frac{\rho -2 \rho_d}{4}\left(\nabla(\theta_1-\theta_2)\right)^2
%\bigg\}.
\end{align}
\par
In the absence of rotation, the model Eq. \eqref{compmodel} has  three different phases.
(i) At low drag and low temperature, there is a phase with broken $U(1)\times U(1)$ 
symmetry. (ii) At high temperatures there is a fully symmetric phase.  (iii) At    
$\rho_d>0$, there is a phase with broken $U(1)$ symmetry only in the phase sum. This is the TPS. 
The phase  diagram was studied in the $J$-current representation in \cite{PSF2}, and in terms of 
proliferation of vortex loops in the Villain model in \cite{Dahl}. In this work, we 
address the question  of the physics of this system when it is subjected to rotation. 
To this end, we have performed large scale Monte Carlo computations on Eq.~\eqref{compmodel}, 
following the procedures of Ref. \onlinecite{Dahl}. Rotation is accounted for by choosing 
$\vect{\Theta}=(0,2\pi fx,0)$, where $f$ is the number of rotation induced vortices per 
plaquette in the $xy$-plane. Our free energy is a function of the ratios of stiffnesses to 
temperature. Thus, we explore the phase diagram in terms of these dimensionless ratios, i.e. 
by absorbing the temperature in $\rho$ and $ \rho_d$. Thus, low values of  $\rho$ and $\rho_d$ amount to 
high temperatures, and vice versa. In these standard units, and in the discretization scheme 
using the Villain-approximation, the single-component $XY$ model has a critical stiffness 
$\rho_c \approx 0.33$. We have considered cubic numerical grids, with periodic boundary 
conditions, of size $L^3$ with $L=64$ and $128$.  Moreover, we have chosen $f=1/64$. For 
each coupling we have used $5\cdot10^5$ sweeps over the entire grid for thermalization, 
and then used $10^6$ sweeps for calculating averages. 
\par
In the limit $\rho_d \to 0$, the system tends towards two decoupled superfluids for which our 
simulations recover the standard triangular vortex lattice forming in response to rotation. 
Already for a drag $\rho_d$ as low as $\rho_d \approx 0.08$, the energetically most favourable 
vortex ordering becomes square vortex lattices for each of the components. These lattices are 
shifted with respect to each other half a lattice spacing in the $x$- and $y$- directions. This 
effect, arising in this hydrodynamic model with current-current interactions, has a counterpart 
in the appearance of square lattices in two-component condensates with density-density interaction 
\cite{Mueller-Ho}. Below, we study vortex matter with further increased drag and temperature  
by (i)  inspection of $3D$ snapshots of typical vortex configuration, (ii) calculating structure 
factors, and (iii) by calculating the quantity $\tilde{\nu}^i(\vect{r}_{\perp})$ representing 
real-space averages over various number of MC sweeps of the vorticity  integrated along the 
$z$-direction, defined as
\begin{equation}
\tilde{\nu}^i(\vect{r}_{\perp})= \left< \frac{1}{L_z}\sum_z\nu_z^i(\vect{r}_{\perp},z) \right>,
\label{ntilde}
\end{equation} 
where $\nu_z^i(\vect{r}_{\perp},z)$ is the vorticity of component $i$ along the $z$ direction at $\vect{r}=(x,y,z)$ 
and $\vect{r}_{\perp}=(x,y)$, $\left<\cdot\right>$ denotes MC averaging. It is important to note  
that when  $\tilde{\nu}^i(\vect{r}_{\perp})$ shows a lattice ordering, although it signals a particular 
rotational response, it does {\it not} necessarily imply a vortex state visible in $z$-axis density 
averages. This is so because  the MC and $z$-axis averages in $\tilde{\nu}^i(\vect{r}_{\perp})$ are 
taken over  vorticity, but not over a density. Vortex segments with opposite phase windings
cancel each others vorticity in these averages.
\par
\begin{figure}[h!!]
 \includegraphics[width=\columnwidth]{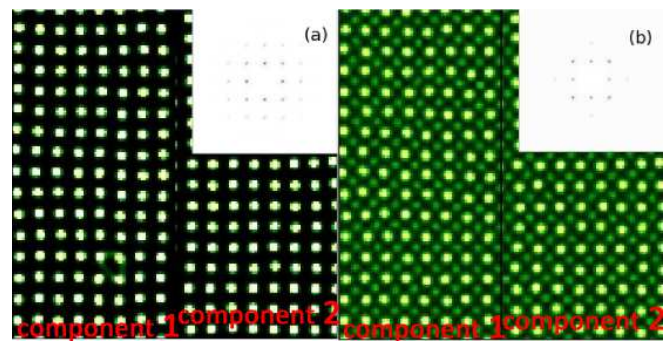}
\caption{(Color online) The average vortex positions  in $xy$-plane integrated along 
the $z$-direction and averaged over every 100th of a total of $1\cdot10^6$ MC sweeps 
($\tilde{\nu}^1(\vect{r}_{\perp})$). Here $\rho=0.924$ and $\rho_d = 0.17$ in a) and 
$\rho_d=0.37$ in b). The brighter green color shows higher probability to find a vortex 
segment directed along the rotation axis.  The left part of each panel shows vortices 
with  phase winding in  component 1 $\tilde{\nu}^1(\vect{r}_{\perp})$, while the 
right part is that for the component 2 $\tilde{\nu}^2(\vect{r}_{\perp})$. The lattice 
of the two components are  displaced a half period in both directions. The inset shows 
the corresponding $\vect{k}$-space structure factor.}
\label{Fig_split_lattice}
\end{figure}

In the low drag, low temperature regime the quantity $\tilde{\nu}^i(\vect{r}_{\perp})$ show peaks corresponding 
to a square lattice. At stronger drag, there appear weak intensity peaks in the center of the plaquettes for 
each of the components, cf. Fig. \ref{Fig_split_lattice}.  This means that an increased drag and temperature 
creates a fluctuating vortex background such that there is an increased probability to find a segment of a vortex 
directed along the $z$-direction and situated in the center of plaquettes of the square vortex lattice,  
Fig. \ref{Fig_split_lattice}. In addition, the higher order Bragg peaks disappear from the $\vect{k}$-space 
structure factor 
$S^{(i)}(\vect{k}_{\perp})~=~\left|\frac{1}{L_xL_yL_zf}\sum_{\vect{r}_{\perp},z}\nu_z^i(\vect{r}_{\perp},z)\ex{-\im \vect{r}_{\perp}\vect{k}_{\perp}}\right |^2$.
\par
\begin{figure}[h!!]
 \includegraphics[width=\columnwidth]{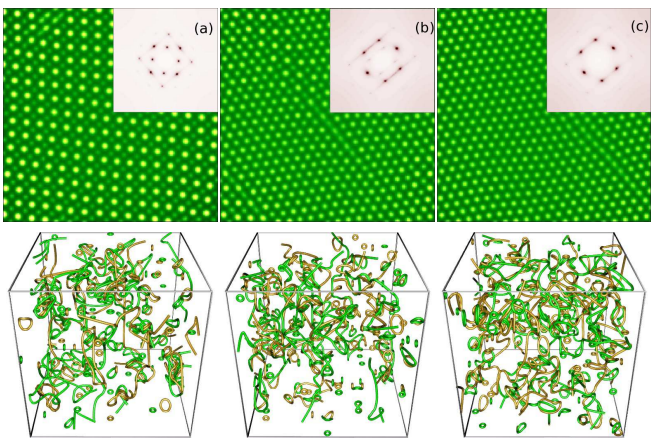}
  \caption{(Color online) The upper row shows the $xy$-position 
  of component 1 vortices integrated along the $z$-direction, $\tilde{\nu}^1(\vect{r}_{\perp})$, 
  averaged over every 100th of a total of $1.0\cdot10^6$ MC sweeps. a): $\rho=0.984$, 
  b): $\rho=0.982$, c): $\rho=0.980$, $\rho_d=0.4$ in all panels. The inset shows the 
  corresponding $\vect{k}$-space structure factor $S^1(\vect{k}_{\perp})$. The bottom row 
  shows typical $3D$-snapshots of the vortex configuration of a $16\times 16\times 16$ 
  segment of the simulated system.  The green  and yellow color represents vortices in 
  different components.
 The figure shows the
  transition from a square lattice structure (leftmost column) to a
  triangular lattice structure (rightmost column). The middle column
  shows coexistence of a triangular and square lattice. The square lattice
is seen in the left top and bottom corner of the panel b), it is also possible to see
a square structure inside the hexagonal structure in the $\vect{k}$-space inset.
For visualizing the $3D$ snapshots, the vortex diameter is chosen  to be $0.1$ of the numerical grid
spacing. Sharp bends arising at the scales of numerical grid spacing are smoothed
by spline interpolation.  For animations, see \cite{link}.}
\label{Fig_square_triang}
\end{figure}

 
With increasing temperature, equivalently decreasing $\rho$, the intensity of new peaks in the quantity  
$\tilde{\nu}^i(\vect{r}_{\perp})$ grows at sufficiently large $\rho_d$. Eventually, we observe a 
discontinuous phase transition to a state with domains where the secondary real-space vortex position 
peaks have an intensity equal to the primary peaks. In these domains, the lattice symmetry also changes 
from square to triangular, and the real space position averages for each component 
become identical, i.e. $\tilde{\nu}^1(\vect{r}_{\perp}) \approx \tilde{\nu}^2(\vect{r}_{\perp})$. 
These triangular lattice domains  coexist with  domains of square high-intensity lattices  with a 
weaker intensity square sublattice, where $\tilde{\nu}^1(\vect{r}_{\perp})$ is approximately the 
same as $\tilde{\nu}^2(\vect{r}_{\perp})$, but shifted a half lattice spacing  in the $x$- and 
$y$-direction. This is seen by comparing panels a) and b) in Fig. \ref{Fig_square_triang}.
\par
\par
\begin{figure}[h!!]
  \includegraphics[width=\columnwidth]{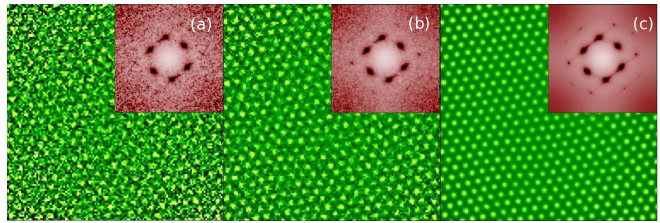} 
 \caption{(Color online) Panel (a) is a typical snapshot of 
$\tilde{\nu}^1(\vect{r}_\perp)$  with the corresponding $\vect{k}$-space structure factor 
as inset, while panel (b) is the corresponding quantity averaged over 5 different configurations 
(100 MC sweeps are used to obtain a new configuration) while in panels (c) the quantity is 
averaged over 1000 different configurations. We see that one needs to average over several 
configurations before a triangular lattice is  clearly visible in $\tilde{\nu}^i$. The computations 
were done for $\rho=0.98$, $\rho_d=0.4$. }
\label{Fig_struc_triang}
\end{figure}

With further increase of temperature, the triangular lattice domains grow until 
$\tilde{\nu}^1(\vect{r}_{\perp})$  and $\tilde{\nu}^2 (\vect{r}_{\perp})$ form identical 
essentially perfect triangular lattices which precisely coincide in space, cf. panels c)
of Fig. \ref{Fig_struc_triang}. 
Note also that now $\tilde{\nu}^i(\vect{r}_{\perp})$ has twice as high density  of vortex positions 
as the low-drag, low-temperature case. Fig. \ref{Fig_square_triang} also shows 
snapshots of typical vortex configurations arising in these states. Even in the state with 
$U(1)\times U(1)$ symmetry and square lattice with relatively weak sublattice intensity 
peaks in $\tilde{\nu}^i(\vect{r}_{\perp})$, it is not obvious from a typical snapshot  
that the system has a vortex lattice with translational symmetry is broken. Moreover, 
it is  impossible to see vortex lattices in any snapshots corresponding to the case when 
statistical averaging produces nearly perfect triangular lattice. Thus, we will use the term 
``hidden vortex lattice" (HVL) for this case. In Fig. \ref{Fig_struc_triang}, we show  
$\tilde{\nu}^i(\vect{r}_{\perp})$ averaged over  a different number of MC snapshots. In the 
case of triangular HVL, in contrast to the low-drag low-temperature case, averaging over a 
small number of MC snapshots shows no vortex lattice in real space images.  As seen from 
$3D$ snapshots, the effect is stronger for density averages, i.e.  averages performed over 
vortex-core positions  instead of vorticity average of $z$-axis directed vortex and antivortex 
segments,  represented by $\tilde{\nu}^i(\vect{r}_{\perp})$. 

\par

 \begin{figure}[h!!]
 \includegraphics[scale=0.6]{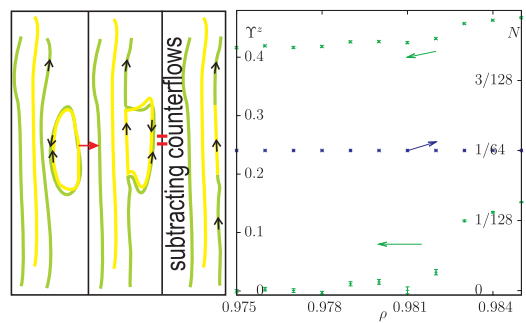}
\label{Fig_PD}
\caption{(Color online)
Left panel illustrates how a segment of a rotation induced vortex line effectively can 
change color via merger with a thermally excited composite $(1,-1)$ vortex loop. The 
process is responsible for e.g. appearance of sublattice peaks in  $\tilde{\nu}^i(\vect{r}_\perp)$. 
Right panel shows that when the system undergoes a transition 
from square to triangular vortex lattice, the helicity modulus for the $(\theta_1 - \theta_2)$ 
sector, $\Upsilon_{-}$ goes to zero, while the helicity modulus for component 1 , $\Upsilon_1$ 
stays finite (left axis, squares).  Also, it is seen that in spite of vortex number doubling in the 
quantity $\tilde{\nu}^i(\vect{r}_\perp)$, the number of $z$-directed rotation induced 
vortex segments is constant (right axis, circles). $N$ is the number density of 
$z$-directed vortices.}
\label{schematic}
\end{figure}

The evolution of vortex matter with increasing temperature at significantly strong drag 
can be described as follows. Topologically, the increase in drag makes composite defects 
$(1,-1)$ with energy $\sim\rho-2\rho_d$ the easiest objects to proliferate. The thermally 
generated composite $(1,-1)$ vortex loops interact with the rotation induced lattice 
through processes schematically illustrated in Fig. \ref{schematic}. For instance, 
a rotation induced $(0,1)$ vortex line can absorb a segment of a thermally created $(1,-1)$ vortex 
loop. This changes the ``color" of a segment of the rotation induced line. Subtracting  
all counterflow segments (i.e. co-centered counter-directed vortices $(1,-1)$) which are not 
directly relevant to rotational response, shows that this process leads to a rotation-induced 
vortex lattice where the vortex lines comprising the lattice will have randomly alternating and 
thermally fluctuating  colors. Indeed, when the number of thermally induced $(1,-1)$ loops is 
low, each sublattice at any moment acquires only a small number of segments of vortices of 
another color via rare merger processes with thermally excited composite vortices. This is the 
origin of the weak sublattice intensity peaks appearing in the centers of each plaquette, 
discussed above. At sufficiently strong drag the  system should undergo a transition to the 
TPS via proliferation of $(1,-1)$ vortices. In the TPS, the remaining broken $U(1)$ symmetry in 
the phase sum still allows the system to form rotation induced lattices of individual vortices, 
but the individual vortices  $(1,0)$ and $(0,1)$ will constantly absorb and emit $(1,-1)$ 
vortex loops. Thus, one cannot attribute a specific color to them. In this state the system 
only has one type of color-indefinite topological defect, and the spatially and MC averaged 
images $\tilde{\nu}^i(\vect{r}_\perp)$  display a triangular lattice.  Even though  the averaged 
real space images show a doubled number of vortices as seen in Fig.  \ref{Fig_struc_triang}, 
Fig. \ref{schematic} illustrates that in our simulations the total number of $z$-components of 
the  elementary segments of rotation-induced vortices does not change. In snapshots, every 
rotation induced vortex line on average consists of 50$\%$ green segments and $50\%$ yellow 
segments. Overall, the system in this state is in a disordered vortex tangle state which is  
continuously undergoing merger processes between composite and individual vortices. The 
breakdown of spatial symmetry transpires only after spatial and MC averaging. 
\par
We now discuss quantitatively the influence of rotation, for given vortex density, 
on the phase diagram of the system. Fig. \ref{Fig_PD} shows the phase diagrams of the  
model  in the Villain approximation, both  without rotation \cite{Dahl}, and with 
rotation. In the zero-drag limit the broken symmetry domain shrinks most significantly, 
since under rotation the symmetry is now restored by lattice melting rather than vortex 
loop proliferation. On the other hand, in the strong-drag limit an opposite situation 
arises. Namely, the transition from $U(1)\times U(1)$ to $U(1)$ TPS state is governed by 
composite vortices and for strong enough drag is almost unaffected by rotation-induced 
lattice of individual vortices. However, the transition  from  TPS to a fully symmetric 
state is  strongly affected by rotation because it is dominated by vortex-lattice melting 
rather than vortex-loop proliferation. Note also that the stiffness at this 
transition is independent of $\rho_d$ and is exactly twice the critical stiffness of the 
phase transition in the zero-drag limit. This demonstrates the accuracy of the  separation 
of variables argument in Ref. \onlinecite{Dahl} in case of a rotating system.

\begin{figure}[h!!]
  \includegraphics[width=\columnwidth]{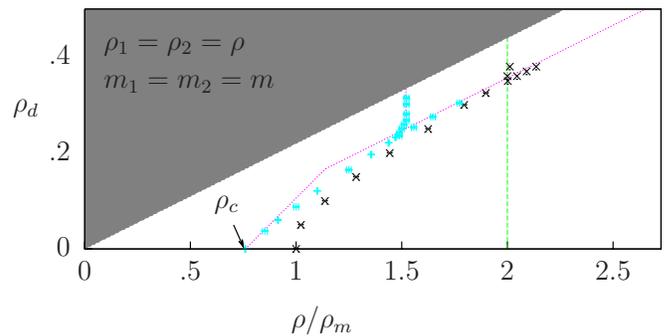}
\caption{(Color online) Phase diagram of the model Eq. \ref{model}
with and without rotation. Dotted lines and blue crosses are obtained from  
the analytical estimates and numerical results from Ref. \onlinecite{Dahl}. 
Black crosses are obtained in the current case of rotating system from the 
vanishing of the peaks of structure function at primary reciprocal lattice 
vectors of the rotation-induced vortex lattice, which signals symmetry 
restoration.}
\label{Fig_PD}
\end{figure}
 
In conclusion, we have studied the rotational response of a hydrodynamic model of two interacting 
superfluids. At very low temperatures, the drag effect results in formation of vortex lattices with 
square symmetry in response to rotation. At moderate temperatures, there appears a statistical vorticity 
buildup in the form of a weak square sublattice. With further elevation of the temperature the system 
undergoes a transition to a TPS which we find is accompanied by a melting of the square lattice into 
a triangular one. The new triangular lattice breaks translation symmetry in a statistical sense. Snapshots 
of this state reveals a highly entangled vortex state. We stress that the quantities which  we use, namely
 $\tilde{\nu}^i(\vect{r}_\perp)$ and $S^i({\vect{k}})$, measure averaged vorticity  and  do not necessarily 
 imply detectable breakdown of translation symmetry in density measurements. This might have experimental 
 implications. There might be regimes where density snapshots may not 
display vortex lattices in the TPS, even though the system may have perfectly triangular HVL from the 
point of view of the above quantities which measure averaged vorticity. It might, however, be possible 
in principle to detect HVL in  interference experiments. Furthermore, in a certain sense a counterpart 
of some of the phenomena discussed above may be  visible in the density profile of quasi-$2D$ systems. 
There, in the regime of strong drag, the  composite $(1,-1)$ vortices undergo a Berezinskii-Kosterlitz-Thouless 
transition at $(\rho/2 -\rho_d)=\pi/4$ while the system retains the order in the phase sum for $\rho/2>\pi/2$.   
Under rotation, they  can be expected to display a vortex lattice of individual vortices coexisting with a liquid 
state of thermally excited composite vortices and antivortices.  Finally, we remark that 
the previous studies of the  effect of a presence of a trap on thermally fluctuating vortices in 
single-component hydrodynamic model \cite{trap}, suggest that a density variation in a trapped 
two-component condensate may produce a  situation where several of the above states may be 
simultaneously present at different distances from the center of the trap.  
\par 
This work was supported by the Norwegian Research Council Grants No. 158518/431 and No. 158547/431 (NANOMAT), 
and Grant No. 167498/V30 (STORFORSK).

\end{document}